\documentclass[prb,letterpaper,twocolumn,showpacs,superscriptaddress,amsmath,amsfonts,amssymb,floatfix,preprintnumbers,citeautoscript]
{revtex4}
\pdfoutput=1
\usepackage{dcolumn,graphicx,color,booktabs} \graphicspath{{Images/}}
\usepackage{amsmath,bm}
\usepackage[outercaption]{sidecap}
\renewcommand{\figurename}{Fig.}
\makeatletter\renewcommand{\fnum@figure}[1]{\figurename~\thefigure~(color online).}\makeatother
\definecolor{DarkBlue}{rgb}{0,0,0.5}

\begin{document} \pagestyle{plain}

\title{Momentum dependence of the superconducting gap in Ba$_{1-x}$K$_{x}$Fe$_2$As$_2$}

\author{D.\,V.~Evtushinsky}
\affiliation{Institute for Solid State Research, IFW Dresden, P.\,O.\,Box 270116, D-01171 Dresden, Germany}
\author{D.\,S.~Inosov}
\affiliation{Institute for Solid State Research, IFW Dresden, P.\,O.\,Box 270116, D-01171 Dresden, Germany}
\affiliation{Max-Planck-Institute for Solid State Research, Heisenbergstrasse 1, D-70569 Stuttgart, Germany}
\author{V.~B.~Zabolotnyy}\author{A.\,Koitzsch}
\affiliation{Institute for Solid State Research, IFW Dresden, P.\,O.\,Box 270116, D-01171 Dresden, Germany}
\author{M.~Knupfer}\author{B.~B\"{u}chner}
\affiliation{Institute for Solid State Research, IFW Dresden, P.\,O.\,Box 270116, D-01171 Dresden, Germany}
\author{M.\,S.\,Viazovska}
\affiliation{Max-Planck-Institute for Mathematics,Vivatsgasse 7, 53111 Bonn, Germany}
\author{G.\,L.\,Sun} \author{V.~Hinkov}
\affiliation{Max-Planck-Institute for Solid State Research, Heisenbergstrasse 1, D-70569 Stuttgart, Germany}
\author{A.\,V.~Boris}
\affiliation{Max-Planck-Institute for Solid State Research, Heisenbergstrasse 1, D-70569 Stuttgart, Germany}
\affiliation{Department of Physics, Loughborough University, Loughborough, LE11 3TU, United Kingdom}
 \author{C.\,T.~Lin} \author{B.\,Keimer}
 \affiliation{Max-Planck-Institute for Solid State Research, Heisenbergstrasse 1, D-70569 Stuttgart, Germany}
\author{A.~Varykhalov}
\affiliation{BESSY GmbH, Albert-Einstein-Strasse 15, 12489 Berlin, Germany}
\author{A.\,A.~Kordyuk}
\affiliation{Institute for Solid State Research, IFW Dresden, P.\,O.\,Box 270116, D-01171 Dresden, Germany}
\affiliation{Institute of Metal Physics of National Academy of Sciences of Ukraine, 03142 Kyiv, Ukraine}
\author{S.\,V.~Borisenko}
\affiliation{Institute for Solid State Research, IFW Dresden, P.\,O.\,Box 270116, D-01171 Dresden, Germany}

\begin{abstract}
\noindent The precise momentum dependence of the superconducting gap in the iron-arsenide superconductor with $T_{\rm c}=32$\,K (BKFA) was determined from angle-resolved photoemission spectroscopy (ARPES) via fitting the distribution of the quasiparticle density to a model. The model incorporates finite lifetime and experimental resolution effects, as well as accounts for peculiarities of BKFA electronic structure. We have found that the value of the superconducting gap is practically the same for the inner $\Gamma$-barrel, X-pocket, and ``blade''-pocket, and equals 9\,meV, while the gap on the outer $\Gamma$-barrel is estimated to be less than 4\,meV, resulting in $2\Delta/k_{\rm B}T_{\rm c}=6.8$ for the large gap, and $2\Delta/k_{\rm B}T_{\rm c}<3$ for the small gap. A large ($77\pm3$\%) non-superconducting component in the photoemission signal is observed below $T_{\rm c}$. Details of gap extraction from ARPES data are discussed in Appendix.
\end{abstract}


\maketitle
\section{Introduction}
\noindent Recently a new class of high-temperature superconductors, iron-based pnictides, attracted much attention due to a rapid increase of the critical temperature, $T_{\rm c}$, up to 56\,K \cite{Wang}. These novel materials, still remaining \textit{terra incognita} for theoreticians and experimentalists, require vast efforts from both sides to achieve a progress in the understanding of their nature. One of the most important contributions that experimentalists can make to the development of a theory of any class of superconductors, is revealing the magnitude and symmetry of the superconducting gap. Knowledge of the precise momentum dependence of the superconducting gap can provide desirable information about the pairing mechanism that underlies superconductivity in these compounds. Up to now there are a number of papers, providing different estimates of the superconducting gap in iron-based superconductors \cite{Chen, Szabo, Mu, Kaminski1, Ding, Zhou, Kaminski, Hasan}, as well as different conclusions about the strength of coupling and applicability of BCS (Bardeen-Cooper-Schrieffer) theory to these compounds. Here we present an angle-resolved photoemission spectroscopy (ARPES) study of the superconducting gap in single crystals of Ba$_{1-x}$K$_{x}$Fe$_2$As$_2$, $T_{\rm c}$=32\,K (BKFA). The superconducting gap is extracted from photoemission data via a fit to a model, that accounts for finite self-energy, temperature, experimental resolution, as well as nonlinearity of the band dispersion, where it is necessary.

\begin{figure}[!h]
\vspace{-3ex}\includegraphics[width=1.0\columnwidth]{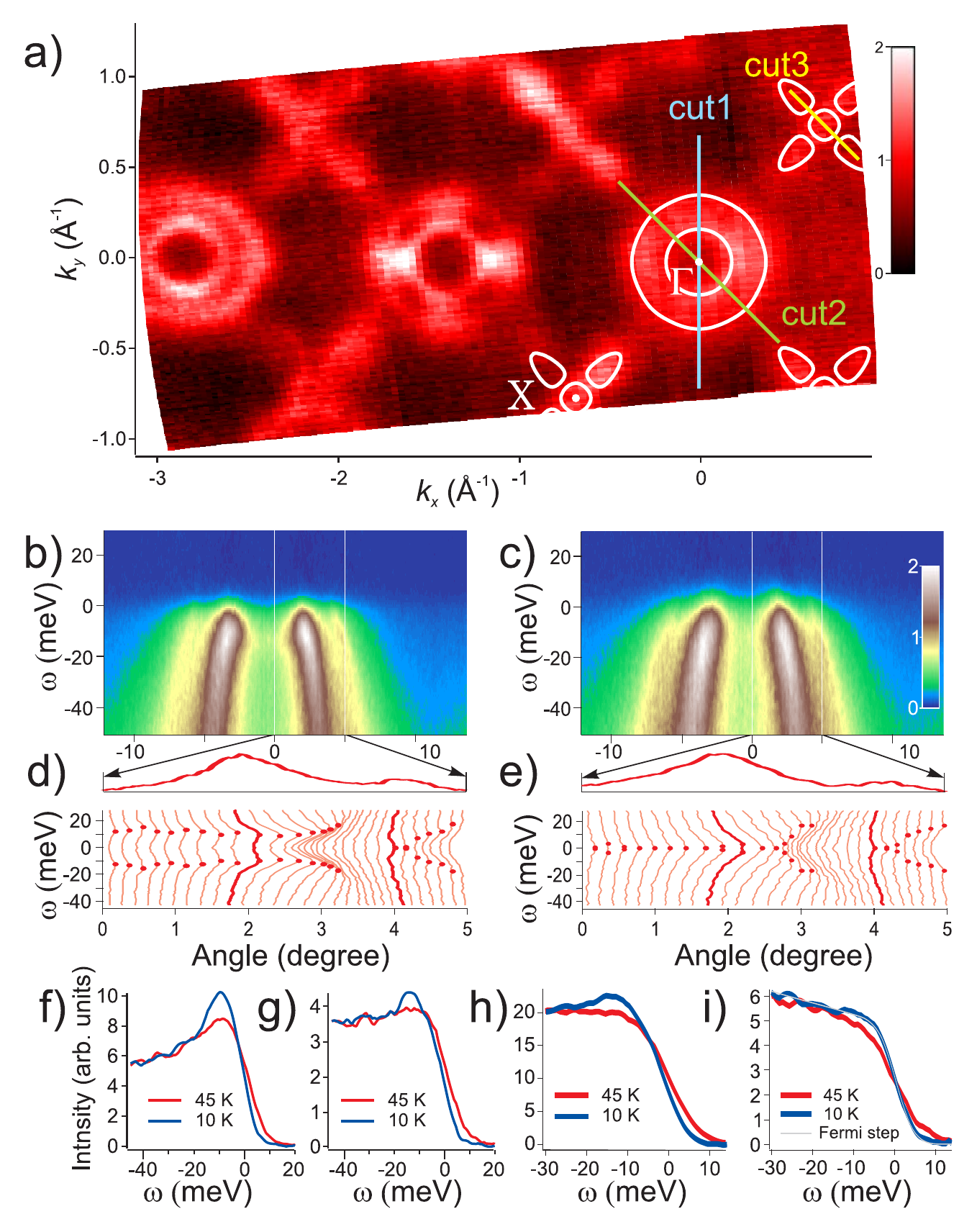}\vspace{-2ex}\caption{(a) Distribution of the photoemission intensity at the Fermi level (FL) with superimposed Fermi surface (FS) contours (white lines). (b) Momentum-energy cut through the $\Gamma$-point [cut1 in panel (a)] taken at 10\,K. (c) Same cut, taken at 45\,K. (d), (e) MDC, taken at the FL, and symmetrized EDC from cuts (b) and (c) respectively. Maxima of the symmetrized EDC are marked by dots. (f) $k_{\rm F}$ EDC referring to the inner $\Gamma$-barrel, recorded at 10 and 45\,K. (g) Near-$k_{\rm F}$ EDC emphasizes onset of the superconductivity even better. (h) Energy dependence of the inner $\Gamma$-barrel intensity, extracted from the fit of MDC. (i) The same for the outer $\Gamma$-barrel.
}
\vspace{-12ex}
\label{f:Model1}
\end{figure}

\begin{figure}[h]
\includegraphics[width=1.0\columnwidth]{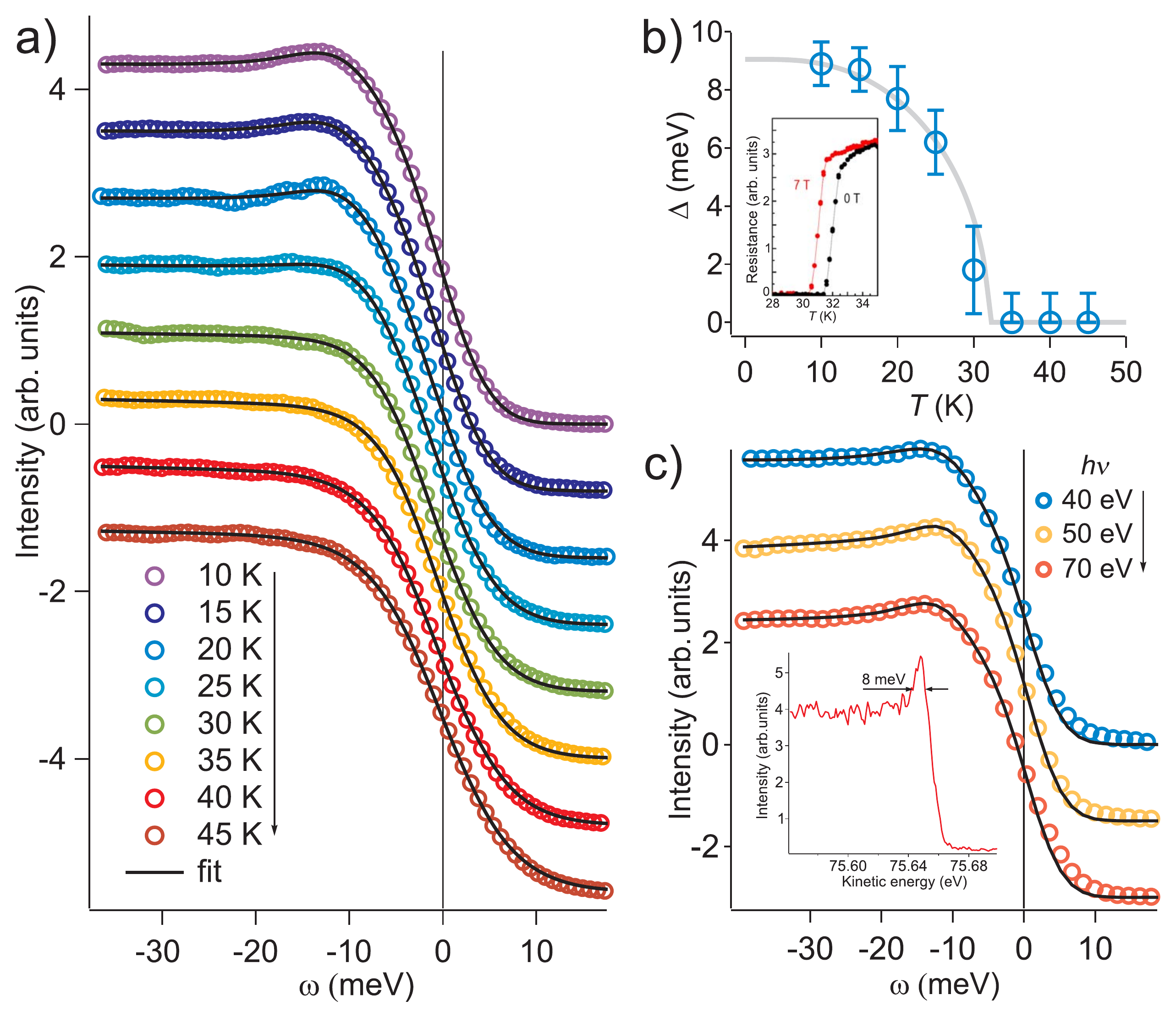}\caption{Temperature and momentum dependence of the superconducting gap on the inner $\Gamma$-barrel. Reliability of the results. (a) Evolution of the integrated EDC from cut1 [see Fig.~1(a)] with temperature and fits to formula (1). EDC are shifted along left axis for clarity. Panel (b) shows the extracted magnitude of the gap, plotted against temperature. Inset to (b) shows temperature dependence of resistivity (with and without magnetic field), confirming high quality of the crystals and emphasizing equality of bulk and surface $T_{\rm c}=32$\,K. (c) Integrated EDC from cut2 [see Fig.~1(a)], measured at different excitation energy at 11\,K, and corresponding fits to formula (1) reveal reproducibility of the data and robustness of the fitting procedure. EDC are shifted along left axis. The inset in (c) shows a single EDC recorded with $h\nu$=80\,eV, demonstrating high resolution at high excitation energies.}
\vspace{-0ex}
\label{f:Model2}
\end{figure}

\section{Results}
According to our recent study \cite{Volodya}, the Fermi surface (FS) of BKFA, as seen in ARPES, consists of four different sheets: outer $\Gamma$-barrel, inner $\Gamma$-barrel, X-pocket, and ``blade''-pockets along the X$\Gamma$ line \cite{blade} [see Fig.~1(a)]. X-pocket is electron-like, while all other FS sheets are hole-like. Fig.~1(b) and (c) show the same energy-momentum cut [cut1 in Fig.~1(a)] through the distribution of the photoemission intensity at 10 and 45\,K respectively. A backfolding dispersion of the inner $\Gamma$-barrel develops with cooling below $T_{\rm c}$ [Fig.~1(b),\,(c)], that points to the opening of the superconducting gap. To investigate the behavior of the quasiparticle density near the Fermi level (FL) in detail, we plot symmetrized energy distribution curves (EDC) measured at 10 and 45\,K in panels (d) and (e) respectively. The distance between the two peaks in the symmetrized EDC approximately equals to the doubled value of the energy gap, $2\Delta$. As follows from Fig.~1(d) and (e), peaks in EDC, which correspond to the FL crossing of inner $\Gamma$-barrel, split into two below $T_{\rm c}$, indicating the opening of a gap of the order of $9$\,meV, while peaks in the EDC, which refer to the outer $\Gamma$-barrel do not split upon cooling, indicating zero (or small in comparison with the peak width) magnitude of the gap on this part of the FS. Fig.~1(f), (g) show the energy dependence of the intensity [area under the momentum distribution curve (MDC)], which comes from inner and outer $\Gamma$-barrels, as extracted from the fit of MDC to four Lorentzians. A pile-up peak clearly develops on the curve that corresponds to the inner $\Gamma$-barrel, while no such feature is observed for the outer $\Gamma$-barrel. The resolution-broadened 10\,K Fermi cut-off is plotted in the panel (g) to show that the difference between 45\,K- and 10\,K-curves mainly comes from temperature smearing of the Fermi edge. The mentioned arguments allow us to conclude that the inner $\Gamma$-barrel bears a gap of the order of $9$\,meV, and the gap on the outer one is much smaller.

Though straightforward and unpretentious, ``symmetrization'' is a rough method for the gap extraction from photoemission data, therefore below we improve the assessment of the gap magnitude with a robust fitting procedure, where the value of the superconducting gap is extracted from the fit of EDC, integrated in a finite momentum window. In this case the integration is performed over a very small, compared to the Brillouin zone size, region, which \emph{does not} imply reduction to momentum-integrated data, and is used only in order to collect the whole available photoemission signal, referring to the particular FL crossing of a single band. The integrated EDC (IEDC) is fitted to the specially derived formula (see Appendix I), which coincides with Dynes function \cite{Dynes} multiplied by the Fermi function, and convolved with the response function:
\begin{equation}
\text{IEDC}(\omega)=\Biggl[f(\omega, T)\cdot \Bigl|\text{Re}\frac{\omega+i\Sigma^{\prime\prime}}{E}\Bigr|\Biggr]\otimes R_{\omega}(\delta E),
\end{equation}
where $E=\sqrt{(\omega+i\Sigma^{\prime\prime})^2-\Delta_{\mathbf{k}}^2}$, $\omega$ is the binding energy with reversed sign, $T$ is the temperature, $\Sigma^{\prime\prime}$ is the imaginary part of the self-energy, $\Delta_{\mathbf{k}}$ is the momentum-dependent superconducting gap, and $\delta E$ is the experimental resolution. A similar method of gap extraction is widely used in angle-integrated photoemission spectroscopy \cite{Kiss}.
Fig.~2(a) shows IEDC, that refer to the inner $\Gamma$-barrel, from cut1 [see Fig.~1(a)], measured with 50\,eV photon energy at different temperatures, as well as their fits to formula (1). The temperature dependence of the extracted gap, shown in Fig.~2(b), illustrates that a superconducting gap develops upon cooling through $T_{\rm c}$, and reaches the value of $9.1\pm0.7$\,meV at low temperatures. Fig.~2(c) represents IEDC from cut2 [see Fig.~1(a)] recorded at 11\,K with different incident photon energies, $h\nu$. The data exhibit good reproducibility, and the values of the gap extracted for different $h\nu$ show only a small scattering within the error bars\,---\,fit results in 9.4, 9.5, and 10.2\,meV for $h\nu=40$, 50, and 70\,eV respectively. In order to emphasize the quality of our data recorded at high excitation energies, we show a single EDC recorded with $h\nu$=80\,eV as an inset to Fig.~2(b). Thus, we can conclude that the momentum anisotropy of the superconducting gap on the inner $\Gamma$-barrel is absent within 1.5\,meV. The outer $\Gamma$-barrel is much less intense than the neighboring inner $\Gamma$-barrel, which complicates the analysis. With the same fitting procedure we estimate the gap on the outer $\Gamma$-barrel to be not more than 4\,meV.

Now we turn to the most interesting and problematic region of the BKFA Fermi surface, that was not completely resolved in previous studies of iron-arsenic superconductors\,---\,a propeller-like structure centered at the X-point [see Fig.1~(a)]. Fig.~3(a) and (b) show the same energy-momentum cut through the X-point [cut3 in Fig.~1(a)] above (36\,K) and below (11\,K) $T_{\rm c}$ respectively. \begin{figure}[b!]
\includegraphics[width=\columnwidth]{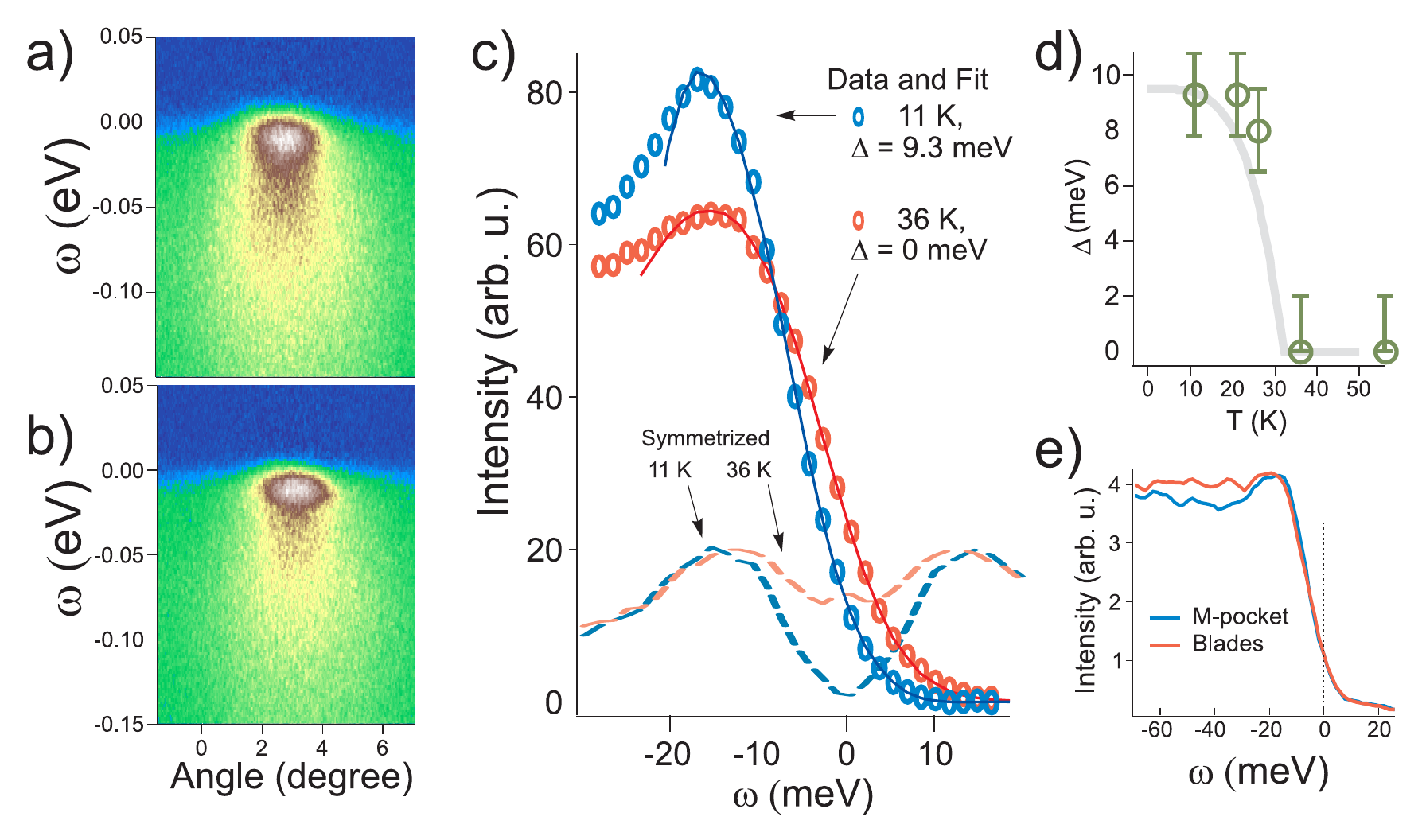} \caption{Superconducting gap on the X-pocket. (a) Energy-momentum cut through the X-point [cut3 on Fig.~1(a)] taken at 36\,K. (b) Same cut taken at 11\,K. (c) Evolution of the IEDC with temperature and fits to formula (2), and symmetrized $k_{\text{F}}$-EDC. (d) Shows temperature dependence of the gap. (e) Comparison of IEDC referring to the M-pocket and to the blades reveals virtually the same values of the superconducting gap.}
\label{f:Model3}
\end{figure}
Note that the intensity of the blades is largely suppressed for this photon energy and light polarization [see Fig.~1(c) in Ref.~\onlinecite{Volodya}]. The difficulties with this region in momentum space are related to the presence of the van Hove singularity close to the FL, which brings the peak in the density of states already above $T_{\rm c}$ [see Fig.~3(c)]. If both bottom and top of the band are far enough from the FL, then one can treat the dispersion of the band as linear without significant accuracy loss, so that formula (1) works well. For the case of band depth comparable to the magnitude of the superconducting gap, formula (1) has to be modified in order to account for the nonlinearity of the normal-state band dispersion. If one assumes that the band possesses electron-like parabolic dispersion with the bottom of the band located at $\omega=-\varepsilon_0$ below the FL, then formula (1) transforms to
\begin{multline}
\text{IEDC}(\omega)=\Biggl[\frac{f(\omega, T)}{2}\cdot \Biggl|\text{Re}\Biggl(\frac{\omega+i\Sigma^{\prime\prime}}{E}\cdot \Biggl[ \sqrt{\frac{\varepsilon_0}{\varepsilon_0-E}}\\+\sqrt{\frac{\varepsilon_0}{\varepsilon_0+E}}\Biggr]+ \sqrt{\frac{\varepsilon_0}{\varepsilon_0-E}}-\sqrt{\frac{\varepsilon_0}{\varepsilon_0+E}}\Biggr)\Biggr|\Biggr]\otimes R_{\omega}(\delta E).
\end{multline}
As it is easy to see, formula (2) reduces to (1) when $\varepsilon_0$ becomes much larger than $\omega$, $\Sigma^{\prime\prime}$, $\delta E$ and $\Delta_{\mathbf{k}}$. For a detailed derivation, see Appendix I.

\begin{figure}[t!]
\vspace{-0ex}\includegraphics[width=1\columnwidth]{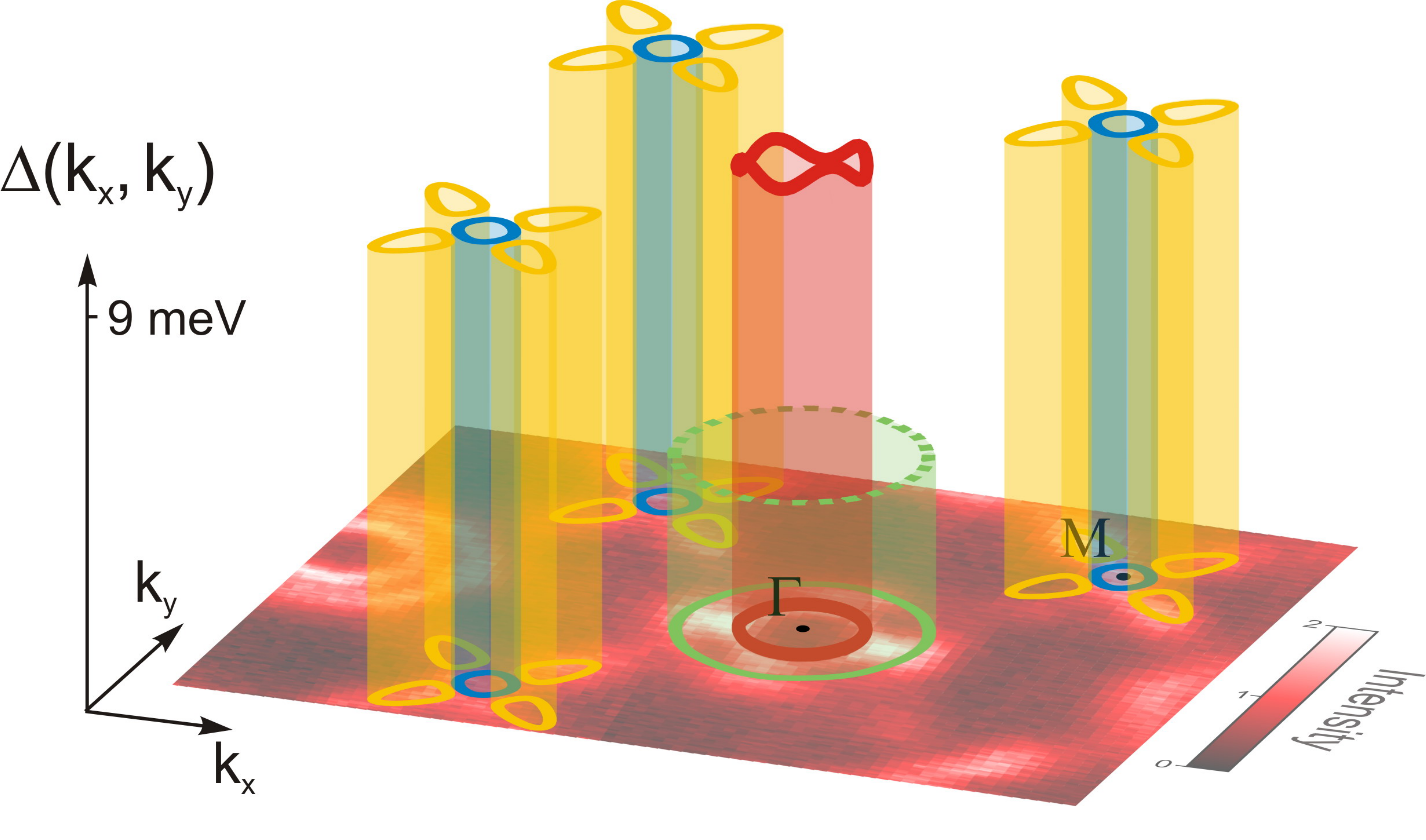} \caption{Momentum dependence of the superconducting gap in Ba$_{1-x}$K$_{x}$Fe$_2$As$_2$ ($T_{\rm c}=32$\,K) is shown as a three-dimensional plot with underlying FS intensity map for orientation. Green line denotes the boundary of the Brillouin zone.}
 \label{f:Model4}
\end{figure}

The depth of the X-pocket was determined from the normal-state data using two different methods\,---\,a fit of momentum distribution curves taken at different binding energies with two Lorentzians, and a fit of the IEDC to formula (2) with $\Delta=0$ and $\varepsilon_0$ as a free parameter. Band depths determined from both methods agree well\,---\,the first method results in $\varepsilon_0=20$\,meV, while the second one yields $\varepsilon_0=20.5$\,meV. Fig.~3(c) shows IEDC from cut3 [see Fig.~1(a)], referring to the X-pocket, measured at 11 and \,36\,K, as well as corresponding fits to formula (2). One may note in Fig.\,3(c) the leading edge below the Fermi level for high temperature data, as well as two separate peaks in symmetrized EDC above $T_{\rm c}$. These signatures of the gap are not relevant here, as discussed in Appendix II. Temperature dependence of the gap, extracted from fitting the data to the formula (2), is shown in the Fig.~3(d). At low temperatures the gap on the X-pocket reaches $9.3$\,meV. From available experimental data we estimate the gap magnitude on the blades to be also 9\,meV [see e.g. Fig.~3(e)]. The results concerning momentum dependence of the superconducting gap are graphically summarized in Fig.~4. The gap is isotropic within the error bars, though, along with similarities to Ref.~\onlinecite{Zhou}, we see evidence for small anisotropy on the inner $\Gamma$-barrel\,---\,the gap may be slightly larger along $\Gamma$X (Brillouin zone diagonal) than along $\Gamma$M (the difference is less than 10\%).

Presented analysis of the data via fitting of IEDC allows us to conclude that the low-temperature spectra have superconducting and non-superconducting components [see Fig.~5]. Only about $23\pm3\%$ of the intensity, coming from the inner $\Gamma$-barrel at 10\,K, refer to the superconducting part of the spectrum \cite{nonSC}. The presence of the two different components in the measured signal can be explained by a phase-separated coexistence of superconducting and normal states, which was already observed in these \cite{Hinkov} as well as in other similar samples \cite{Uemura}.

\begin{figure}[t!]
\vspace{-0ex}\includegraphics[width=0.95\columnwidth]{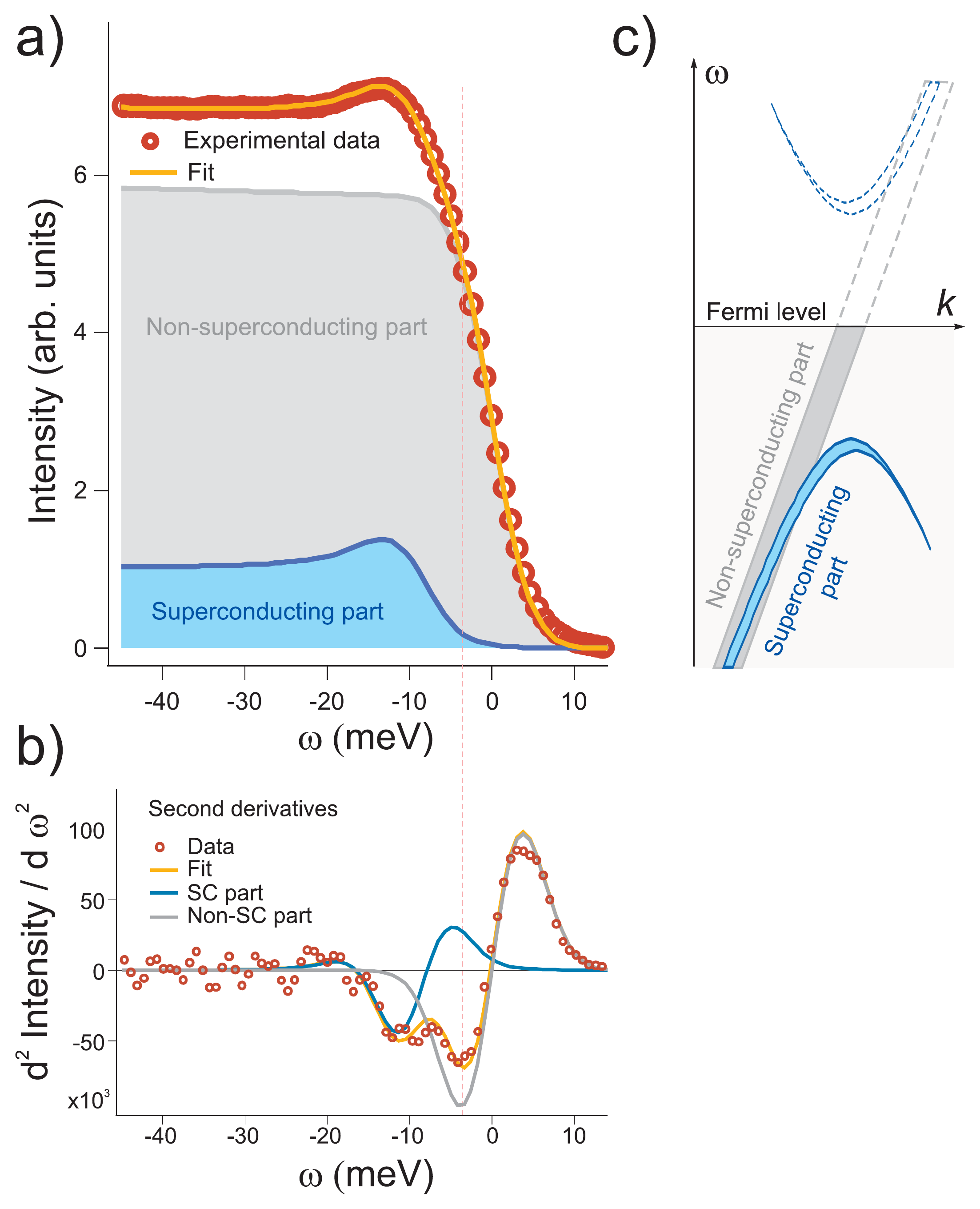} \caption{Superconducting and non-superconducting constituents of the spectrum. (a) Energy distribution of the intensity, corresponding to superconducting and non-superconducting parts of the spectrum. (b) Second derivatives of the data and fit. Structure of the second derivative confirms presence of superconducting and non-superconducting components. (c) Sketch, illustrating presence of two different components in the same spectrum.}
 \label{f:Model4}
\end{figure}

\section{Experimental details}
Single crystals of BKFA were grown using Sn as flux in a zirconia crucible. The crucible was sealed in a quartz ampoule filled with Ar and loaded into a box furnace. A cooling rate of 3\,°C/h was applied from the maximum temperature of 850 down to 550\,°C for the growth. The growth details are described in Ref. \onlinecite{Crystals}. The crystals were cleaved \textit{in situ} and measured with Scienta SES R4000 analyzer at the base pressure of $5\cdot10^{-11}$\,mBar.  ARPES experiments were performed using the ``$1^3$ ARPES'' end station at BESSY.

\section{Conclusions}
In conclusion, we have proposed a precise procedure for extracting the momentum dependence of the superconducting gap from ARPES spectra. The developed method of data treatment allows to measure energy gaps with an accuracy much higher than experimental resolution, similarly to the Voigt-fit procedure \cite{Voigt}, enabling detection of the true values for the MDC width with an accuracy much better than momentum broadening. The IEDC-fitting procedure, applied to ARPES spectra of BKFA, yielded the following results: (i) the gap on the inner $\Gamma$-barrel along $\Gamma$M equals $9.1\pm0.7$\,meV and along $\Gamma$X $9.7\pm1$\,meV; (ii) the gap on the outer $\Gamma$-barrel is less than 4\,meV; (iii) the gap on the X-pocket equals $9.3\pm2$\,meV; (iv) the gap on the blades is estimated to 9\,meV; (v) at 10\,K the imaginary part of the self-energy, $\Sigma^{\prime\prime}$, in the vicinity of the FL was found to be equal to $1$--$2$\,meV. Comparison with other ARPES studies of the superconducting gap in iron-based superconductors is shown in Table~I. We evaluate the coupling strength as $2\Delta/k_{\rm B}T_{\rm c}$=6.8 for the inner $\Gamma$-barrel, X-pocket, and blades, while for the outer $\Gamma$-barrel $2\Delta/k_{\rm B}T_{\rm c}$<3. A comparison to other experiments is shown in Table~II.

\begin{table}[]
\begin{tabular}{l@{~~~}l@{~~}l@{~~}l@{~~}l@{~~}l@{~~}l}
\toprule
Ref. num. &\onlinecite{Kaminski1}& \onlinecite{Ding}& \onlinecite{Zhou}& \onlinecite{Kaminski}& \onlinecite{Hasan}& This paper\\
$\quad \quad \quad \quad \quad T_{\rm c}$& 53\,K & 37\,K & 35\,K & 53\,K & 37\,K & 32\,K\\
\midrule
Inner $\Gamma$-barrel&20&12.5&12&15&12&$9.2\pm1$\\
Outer $\Gamma$-barrel&---&5.5&8&---&6&$<4$\\
X-pocket&---&(12.5)&(10)&---&(11)&$9\pm2$\\
Blades&---&---&(11)&---&---&$9\pm3$\\
Gap anisotropy&---&$<3$&2&$<5$&$<3$&$<1.5$\\
\bottomrule
\vspace{-3ex}
\end{tabular}
\caption{Momentum dependence of the superconducting gap in iron-arsenic superconductors, as revealed by ARPES studies, sorted by the time of appearance on the arXiv.org. Values of the gap and estimates of the gap anisotropy on the inner $\Gamma$-barrel are given in millielectron-volts.}
\label{tab1}
\end{table}
\begin{table}[]
\begin{tabular}{l@{~~~\,}l@{~~}l@{~~}l@{~~}l@{~~}l@{~~}l@{~~}l@{~~}l@{~~}l}
\toprule
Ref. num. &\onlinecite{Kaminski1}&\onlinecite{Ding}& \onlinecite{Zhou}& \onlinecite{Kaminski}& \onlinecite{Hasan} & \onlinecite{Chen}& \onlinecite{Szabo}& \onlinecite{Mu}& This paper\\
\midrule
Large gap& 9 & 8.1 & 8.2 & 6.8 & 7.5 & 3.7 & 9.6 & 4 &   6.8\\
Small gap\,&---&3.6&5.5&---&3.9&---&3.4&---&   $<3$\\
\bottomrule
\vspace{-3ex}
\end{tabular}
\caption{Coupling strength, $2\Delta/k_\text{B}T_\text{c}$, in iron-arsenic superconductors, as revealed by different experimental techniques\,---\,compare to the BSC universal value 3.53. Most of the available studies reveal two superconducting gaps of different magnitudes, which are represented in the table as ``large'' and ``small''. Refs.~\onlinecite{Kaminski1, Ding, Zhou, Kaminski, Hasan} are ARPES studies, Refs.~\onlinecite{Chen, Szabo} are Andreev spectroscopy studies, Ref.~\onlinecite{Mu} is a specific heat study.}
\label{tab1}
\end{table}

Finally, the observation of drastically different superconducting gaps on the inner and outer $\Gamma$-barrels is inline with theoretically suggested magnetic downfolding \cite{Mazin} and with a hidden ($\pi$, $\pi$)-order observed experimentally \cite{Volodya}. Otherwise it would be hard to expect so different gaps on closely located and very similar bands, formed by slightly different combinations of the same atomic orbitals.

\section{Acknowledgements}
The project is part of the FOR538 and was supported by the DFG under Grants No. KN393/4 and BO1912/2-1. We thank I.\,I.\,Mazin, A.\,N.\,Yaresko and M.\,M.\,Korshunov for useful discussions, as well as R.\,H\"{u}bel, R.\,Sch\"{o}nfelder and S.\,Leger for technical support.

\section{Appendix I: Fitting formula}
\subsection{Derivation}
\noindent Below we adduce the detailed derivation of the formulae (1) and (2). We also show that for the simple case of negligible curvature of the band dispersion and momentum independent gap, formula (1) coincides with the Dynes formula.

A very general model for the measured ARPES signal is \cite{kord_LEG, no_matrix_elements}
\begin{equation}
I(k,\omega) =\bigl[f(\omega, T)\,A(k,\omega)\bigr]\otimes R_{\omega}\otimes R_{k}.
\label{signal_model}
\end{equation}
By definition, the integrated EDC is
\begin{equation}
\text{IEDC}(\omega) \equiv \int\!I(k,\omega)\mathrm{d}k.
\label{IEDC_definition}
\end{equation}
As soon as we anyway integrate our data over $k$, momentum resolution does not affect IEDC \cite{int_conv}, which is already an advantage of this method.
Substituting (\ref{signal_model}) into (\ref{IEDC_definition}), we get
\begin{equation}
\text{IEDC}(\omega) = \Biggl[f(\omega, T)\,\biggl(\int\!A(k,\omega)\mathrm{d}k \biggr) \Biggr] \otimes R_{\omega}.
\label{IEDC_formula}
\end{equation}

For the spectral function $A(k,\omega)$ in the superconducting state, we use the following well accepted model \cite{Mahan}:
\begin{equation}
\begin{split}
         A(k,\omega) =
         2 \pi [ u^2_k \delta(\omega - E_k ) +  v^2_k \delta(\omega + E_k) ],
\end{split}
\label{BCSspectralFunction}
\end{equation}
where
    \begin{multline}
    u_k^2 = \frac{1}{2} \left( 1+ \frac{\xi_k}{E_k} \right), \quad
    v_k^2 = \frac{1}{2} \left( 1 -  \frac{\xi_k}{E_k} \right), \quad
    \\E_k=\sqrt{\xi^2_k +\Delta^2}.
    \label{gapequations}
    \end{multline}

\begin{figure}[b]
\vspace{+0ex}\includegraphics[width=0.9\columnwidth]{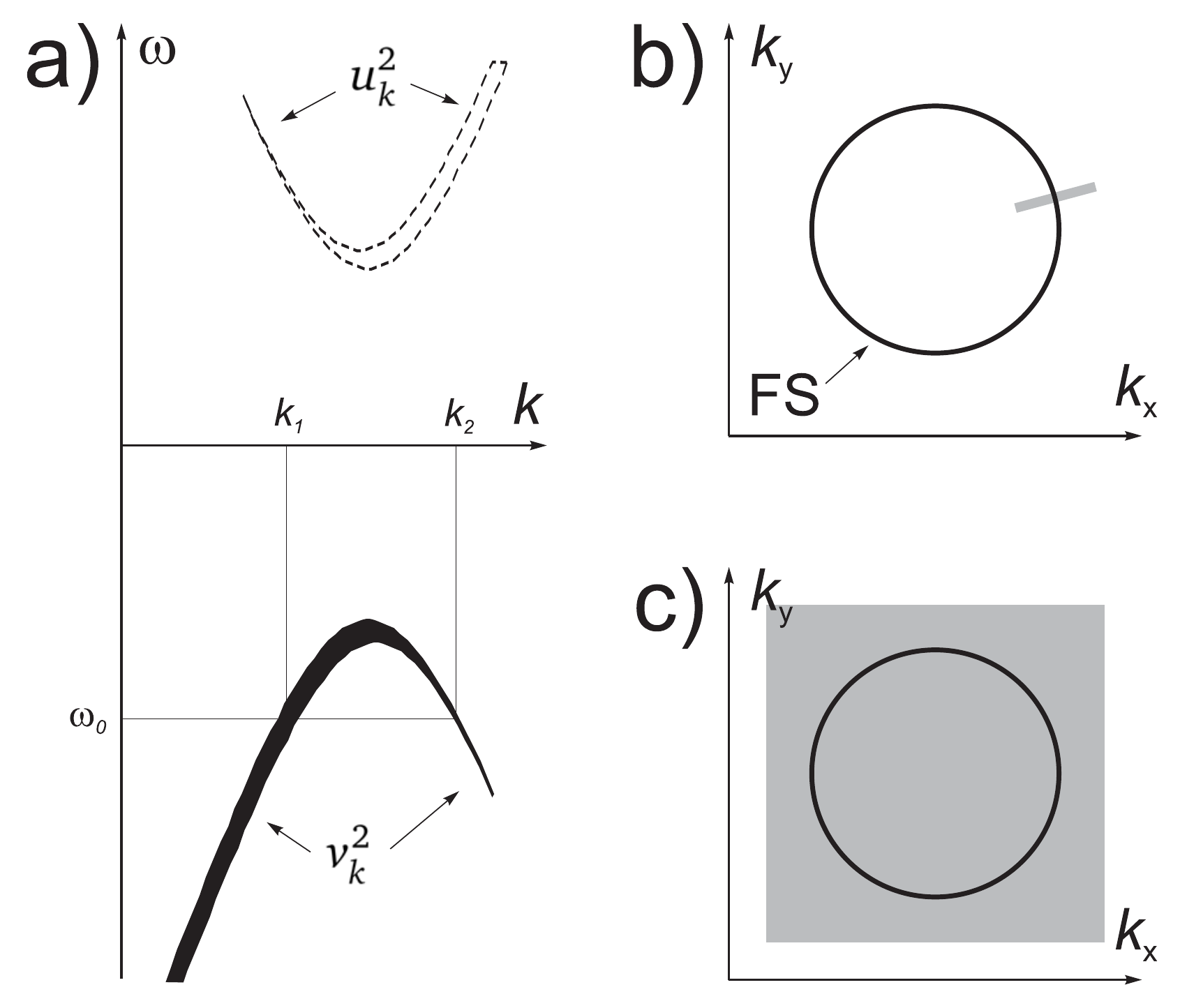}\caption{a) Spectral function in the superconducting state for the case of linear normal-state dispersion $\xi_{k}$. According to the formula (\ref{BCSspectralFunction}), the spectral weight above the Fermi level is governed by $u_k^2$, and by $v_k^2$ below. b) Integration in our case is performed along one energy-momentum cut (grey stroke), which intersects Fermi surface at \emph{only one} point. c) In the momentum-integrated techniques integration is naturally performed over the \emph{whole} momentum space.}
\vspace{-0ex}
\label{f:Model7}
\end{figure}

Substituting (\ref{BCSspectralFunction}) under the integral in (\ref{IEDC_formula}) and omitting unnecessary constant factors, we get
\begin{multline}
\int\!A(k,\omega_0)\mathrm{d}k=\int\!\biggl[u^2_k \delta(\omega_0 - E_k ) +  v^2_k \delta(\omega_0 + E_k )\biggr] \mathrm{d}k=\\
\frac{1}{2}\left( 1- \frac{\sqrt{\omega_0^2-\Delta^2}}{\omega_0} \right)\, \biggl|\frac{\mathrm{d}E_k}{\mathrm{d}k\,} \biggr|^{-1}\biggl|_{k=k_1}+\\ \frac{1}{2}\left( 1+ \frac{\sqrt{\omega_0^2-\Delta^2}}{\omega_0} \right)\,\biggl|\frac{\mathrm{d}E_k} {\mathrm{d}k\,}\biggr|^{-1}\biggl|_{k=k_2},\quad\quad\text{ }
\end{multline}
where $k_{1,2}$ are solutions of $\xi_{k_{1,2}}=\pm\sqrt{\omega_0^2-\Delta^2}$. Below we will denote derivative by a prime:
$\frac{\mathrm{d}E_k} {\mathrm{d}k\,}\bigl|_{k=k_1}\equiv E^{\prime}_{k_1}.$

Expanding the derivative
\begin{equation}
\bigl|E^{\prime}_{k}\bigr| = \frac{\sqrt{\omega_0^2-\Delta^2}}{|\omega_0|}\bigl|\xi^{\prime}_{k}\bigr|,
\end{equation}
we get

\begin{multline}
\int\!A(k,\omega_0)\mathrm{d}k=  \frac{|\omega_0|}{2\sqrt{\omega_0^2-\Delta^2}} \Bigl[\bigl|\xi^{\prime}_{k_1}\bigr|^{-1}+  \bigl|\xi^{\prime}_{k_2}\bigr|^{-1}\Bigr]+\\ \frac{1}{2}\text{sign}(\omega_0)\Bigl[\bigl|\xi^{\prime}_{k_1}\bigr|^{-1} - \bigl|\xi^{\prime}_{k_2}\bigr|^{-1}\Bigr].
\label{narrow_end_formula}
\end{multline}
(In this formula $\xi^{\prime}_{k_{1,2}}$ implicitly depend on $\omega_0$.)

For the case of the linear band dispersion the derivative is constant, $\xi^{\prime}_k=\text{const}$, and we arrive at
\begin{equation}
\int\!A(k,\omega_0)\mathrm{d}k = \frac{|\omega_0|}{\sqrt{\omega_0^2-\Delta^2}}.
\end{equation}

This formula coincides with the Dynes function, although the premises for the latter are somewhat different, requiring the assumption of the momentum independent gap. Important difference in definition of our IEDC and well known Dynes function is that the former is a trace integral along one direction [see Fig.\,6\,(b)], while the latter is a double integral over the whole momentum space [see Fig.\,6\,(c)]:
\begin{equation}
\text{Dynes}(\omega)\equiv\iint\!A\kern1pt(\mathbf{k},\omega)\,\mathrm{d}k_x \mathrm{d}k_y.
\end{equation}
Substituting here the aforementioned model for the spectral function (\ref{BCSspectralFunction}), we go from a double integral to the integration along the contour
\begin{multline}
\text{Dynes}(\omega_0)=\oint\limits_{\mathbf{k}:\xi_\mathbf{k}=\sqrt{\omega_0^2-\Delta^2}} v^2_\mathbf{k}\bigl|\operatorname{\nabla}\!E_\mathbf{k}\bigr|^{-1} \mathrm{d}k \\+ \oint\limits_{\mathbf{k}:\xi_\mathbf{k}=-\sqrt{\omega_0^2-\Delta^2}} v^2_\mathbf{k}\bigl|\operatorname{\nabla}\!E_\mathbf{k}\bigr|^{-1} \mathrm{d}k.
\end{multline}
When the depth of the band is much larger than the value of the superconducting gap, i.e. when we can neglect the nonlinearity of the dispersion (which is an important condition for the Dynes formula to hold!), this expression reduces to the integral over the Fermi surface:
\begin{equation}
\text{Dynes}(\omega_0)=\oint\limits_{\mathbf{k}:\xi_\mathbf{k}=0} \bigl|\operatorname{\nabla}\!E_\mathbf{k}\bigr|^{-1} \mathrm{d}k.
\end{equation}
Here we expand $\operatorname{\nabla}\!E_\mathbf{k}$ similarly to formula (9), and get
\begin{equation}
\text{Dynes}(\omega_0)=\oint\limits_{\mathbf{k}:\xi_\mathbf{k}=0} \frac{|\omega_0|}{\sqrt{\omega_0^2-\Delta^2}} \bigl|\operatorname{\nabla}\!\xi_\mathbf{k}\bigr|^{-1} \mathrm{d}k.
\end{equation}
As soon as $\omega_0$ and $\Delta$ (in this case) do not depend on $\mathbf{k}$, one can pull them out from under the integral:
\begin{equation}
\text{Dynes}(\omega_0)=\frac{|\omega_0|}{\sqrt{\omega_0^2-\Delta^2}} \oint\limits_{\mathbf{k}:\xi_\mathbf{k}=0} \bigl|\operatorname{\nabla}\!\xi_\mathbf{k}\bigr|^{-1} \mathrm{d}k.
\end{equation}
The integrand does not depend on $\omega_0$, therefore the whole integral is an unnecessary for our purposes constant factor, which can be omitted, and we arrive at the same result as (11):
\begin{equation}
\text{Dynes}(\omega_0)=\frac{|\omega_0|}{\sqrt{\omega_0^2-\Delta^2}}.
\end{equation}

\subsection{Finite lifetime}
\noindent Up to now we have the result [formula~(\ref{narrow_end_formula})], obtained under the assumption of infinitely large lifetime, or, in other words, for very sharp bands. In such a case in order to get formula that incorporates effects of the finite lifetime, the following recipe is often used: take the formula, derived for infinite lifetime, add to the argument the imaginary part, and take real part of the result,
\begin{equation}
g(\omega) \rightarrow \text{Re }g(\omega + i\Sigma^{\prime\prime}).
\end{equation}
Below we show that in our case this trick provides the exact result.

In order to account for lifetime broadening rigorously, one has to substitute the delta function in (\ref{BCSspectralFunction}) for a Lorentzian: $$\delta(\omega - E_k) \rightarrow L^{\Sigma^{\prime\prime}}(\omega - E_k) = \frac{1}{2\pi}\frac{\Sigma^{\prime\prime}}{(\omega - E_k)^2+{\Sigma^{\prime\prime}}^2},$$
which results in the possibility to rewrite the expression for the spectral function in the following way:
    \begin{equation}
         A^{\Sigma^{\prime\prime}}(k,\omega) = A(k,\omega)\otimes L^{\Sigma^{\prime\prime}}(\omega - E_k),
    \end{equation}
where $A(k,\omega)$ stands for non-broadened spectral function (\ref{BCSspectralFunction}).
As convolution over $\omega$ commutes with integration over $k$,
\begin{equation}
    \int\!A^{\Sigma^{\prime\prime}}(k,\omega)\mathrm{d}k = \biggl[\int\!A(k,\omega)\mathrm{d}k\biggr]\otimes L^{\Sigma^{\prime\prime}}.
\end{equation}
We already know the result for integration of the spectral function over momentum\,---\,formula~(\ref{narrow_end_formula}), and now the only problem is to evaluate the convolution. We will do it for linear and quadratic band dispersions, i.e. input parameters to derive formulae (1) and (2).

Let $g(\omega)\equiv\int\!A(k,\omega)\mathrm{d}k$, then in order to evaluate the convolution in (20), we have to calculate the integral $\int\limits^{+\infty}_{-\infty}g(\omega) L^{\Sigma^{\prime\prime}}(\omega_0 - \omega)\mathrm{d}\omega.$ $\operatorname*{Res}\limits_{~z=z_0}f(z)=1$

\begin{figure}[]
\vspace{+0ex}\includegraphics[width=0.75\columnwidth]{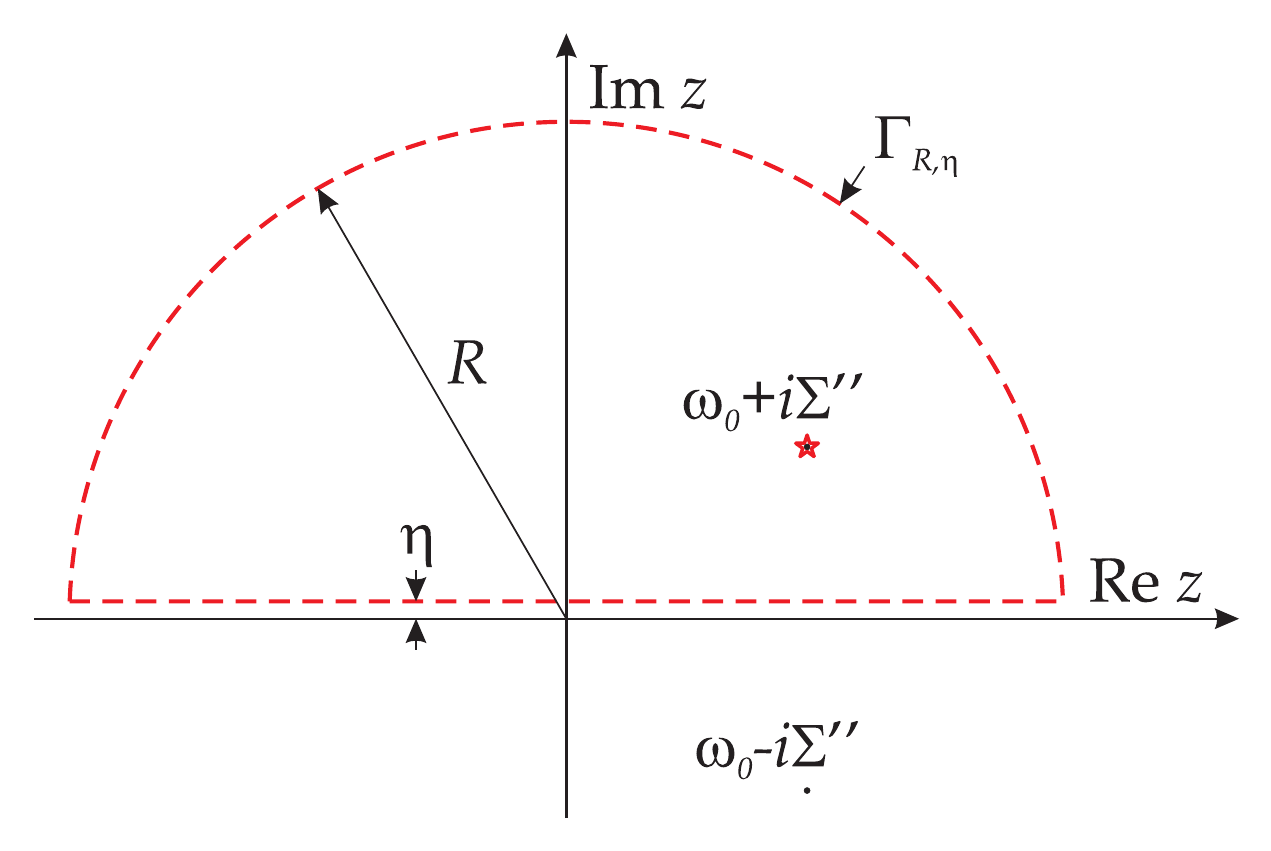}\caption{Intergation along the contour on the complex plane. According to Cauchy's residue theorem, the integral along the contour $\Gamma_{R,\eta}$ equals to the residue in the pole of the integrand inside, $\omega_0+i\Sigma^{\prime\prime}$.}
\vspace{-0ex}
\label{f:Model7}
\end{figure}

The function $g(\omega)$ is defined on the real axis. Once we know the analytic function $\tilde{g}(z)$, $z\in\mathbb{C}$, such that $ \mathrm{Re}(\tilde{g}(\omega))=g(\omega)$ for  $\omega\in\mathbb{R}$, we can calculate the required integral with the help of Cauchy's residue theorem:
\begin{multline}
\int\limits^{-\infty}_{+\infty}g(\omega) L^{\Sigma^{\prime\prime}}(\omega_0 - \omega)\mathrm{d}\omega = \\ \mathrm{Re} \biggl[ \int\limits^{-\infty}_{+\infty}\tilde{g}(\omega) L^{\Sigma^{\prime\prime}}(\omega_0 - \omega)\mathrm{d}\omega\biggr] = \\ \mathrm{Re}\biggl[\lim_{R\rightarrow\infty, \eta\rightarrow0}\oint\limits_{\Gamma_{R,\eta}}\tilde{g}(z) L^{\Sigma^{\prime\prime}}(\omega_0 - z)\mathrm{d}z\biggr] =  \\  \text{\raisebox{-1pt}{\scalebox{1.3}{$\wr$}} $\tilde{g}$ possesses no poles inside $\Gamma_{R,\eta}$ \raisebox{-1pt}{\scalebox{1.3}{$\wr$}}} \\ =\mathrm{Re}\biggr[ 2\pi i \cdot \operatorname*{Res}\limits_{z=\omega_0+i\Sigma^{\prime\prime}}\tilde{g}(z) \frac{1}{2\pi}\frac{\Sigma^{\prime\prime}}{(\omega_0 - z)^2+{\Sigma^{\prime\prime}}^2}\biggl] = \\ \mathrm{Re }\,\,\tilde{g}(\omega_0+i\Sigma^{\prime\prime}),
\end{multline}
which coincides with formula (18), and implies formulae (1) and (2) as corollaries. For definition of the integration contour $\Gamma_{R,\eta}$ refer to Fig.\,7. Explicit form of the function $\tilde{g}(z)$ for linear band dispersion is
\begin{equation}
\tilde{g}_1(z)=\frac{z}{\sqrt[*]{z^2-\Delta^2}},
\end{equation}
where for $z=r e^{i\phi}$ we pick the following definition of the square root $\sqrt[*]{z}\equiv r^{1/2}e^{i\phi/2}$, $\phi\in[0, 2\pi)$.

For quadratic dispersion we get
\begin{multline}
\tilde{g}_2(z)= \frac{1}{2}\frac{z}{\sqrt[*]{z^2-\Delta^2}} \Bigr(\frac{1}{k_1}+\frac{1}{k_2}\Bigl)+\frac{1}{2}\Bigr(\frac{1}{k_1}-\frac{1}{k_2}\Bigl),
\end{multline}
where $k_{1,2}=\sqrt[**]{\varepsilon_0\pm\sqrt[*]{z^2-\Delta^2}}$, $\sqrt[**]{z}\equiv r^{1/2}e^{i\phi/2}$, $\phi\in[-\pi, \pi)$.

Defined in such way, $\tilde{g}_{1,2}(z)$ are analytic in $\mathbb{C}\setminus(-\infty,+\infty)$, i.e. all conditions for Cauchy's residue theorem are fulfilled.

\subsection{Formulae in real numbers}
For numerical calculations it is useful to rewrite formula (1) without the use of complex numbers:
\begin{equation}\tag{$1^{\prime}$}
\text{IEDC}(\omega)=\Biggl[f(\omega, T)\Bigl|\frac{\omega(a+c)+\Sigma^{\prime\prime}b}{\sqrt{2} c \sqrt{a+c}}\Bigr|\Biggr]\otimes R_{\omega}(\delta E),
\end{equation}
where $a=\omega^2-\Sigma^{\prime\prime2}-\Delta_{\mathbf{k}}^2$, $b=2\Sigma^{\prime\prime}\omega$, and $c=\sqrt{a^2+b^2}$.

Similarly, formula (2) can be rewritten as
\begin{multline}\tag{$2^{\prime}$}
\text{IEDC}(\omega)=\Biggl[f(\omega, T)\cdot\frac{1}{2}\sqrt{\varepsilon_0} \cdot \\ \Biggl(\frac{|\omega[(a+c)(\alpha_1+\gamma_1)+b\beta_1]+\Sigma^{\prime\prime}[b(\alpha_1+\gamma_1)-\beta_1(a+c)]|}
{2c\gamma_1\sqrt{a+c}\sqrt{\alpha_1+\gamma_1}}\\
+\frac{|\omega[(a+c)(\alpha_2+\gamma_2)+b\beta_2]+\Sigma^{\prime\prime}[b(\alpha_2+\gamma_2)-\beta_2(a+c)]|}
{2c\gamma_2\sqrt{a+c}\sqrt{\alpha_2+\gamma_2}}\\
-\text{sign}(\omega)\biggl[\frac{\sqrt{\alpha_1+\gamma_1}}{\sqrt{2}\gamma_1}-
\frac{\sqrt{\alpha_2+\gamma_2}}{\sqrt{2}\gamma_2}\biggl]\Biggr)\Biggr]\otimes R_{\omega}(\delta E),
\end{multline}
where
\begin{multline}\nonumber
a=\omega^2-\Sigma^{\prime\prime2}-\Delta_{\mathbf{k}}^2, \quad b=2\Sigma^{\prime\prime}\omega, \quad c=\sqrt{a^2+b^2},\\
\alpha_{1,2}=\varepsilon_0\mp\sqrt{\frac{a+c}{2}}, \quad \beta_{1,2}=\pm\frac{b}{\sqrt{2}\sqrt{a+c}}, \\ \text{and } \gamma_{1,2}=\sqrt{\alpha_{1,2}^2+\beta_{1,2}^2}.
\end{multline}

\section{Appendix II: Extraction of the gap from the modeled data by ``symmetrization'', ``leading edge'', and fitting}

The ``symmetrization'' is highly valued by some part of the ARPES community. We strongly believe that ``symmetrization'' is to be substituted by more rigorous and advanced ways of data treatment, such as those used in a very recent publications on photoemission spectroscopy of superconductors \cite{Valla, Shin}.

Below we model ARPES spectra with formulae (\ref{signal_model}) and (\ref{IEDC_definition}), and extract the gap with ``symmetrization'', ``leading edge'' \cite{kord_LEG}, and proposed here fit of the IEDC. Results confirm that the fitting procedure is robust against momentum resolution, properly accounts for energy resolution and finite lifetime, provides correct values even in the case of the nonlinear band dispersion, and allows one to disentangle non-superconducting and superconducting parts of the spectrum. At the same time, ``symmetrization'' and ``leading edge'' are not stable with respect to the effects of the experimental resolution, and furthermore fail in the case of the shallow band and in the presence of the non-superconducting component.

\subsection{Energy resolution}
First, we study the influence of the experimental energy resolution on the determination of the gap from ARPES data with ``symmetrization'', ``leading edge'', and fit to formula (1) from the Manuscript. The results of these studies are shown in Fig.\,8 and summarized in Table~III. Please note that \emph{not} the resolution of the analyzer is important, but the resolution of the \emph{whole} photoemission experiment. Also it is worthwhile to mention that effects of the lifetime broadening are in some respect similar to the effects of energy resolution, as they both lead to the broadening of the spectra.

By the way, \emph{leading edge} (the lowest binding energy at which the $k_{\rm F}$ EDC reaches half of its maximum) alone is not a good measure of the gap (see corresponding columns in Figs.\,8--10), while \emph{leading edge shift} (shift of the leading edge with respect to the position in the normal state) is a lot more relevant quantity.

\begin{figure*}[]
\vspace{+0ex}\includegraphics[width=1\textwidth]{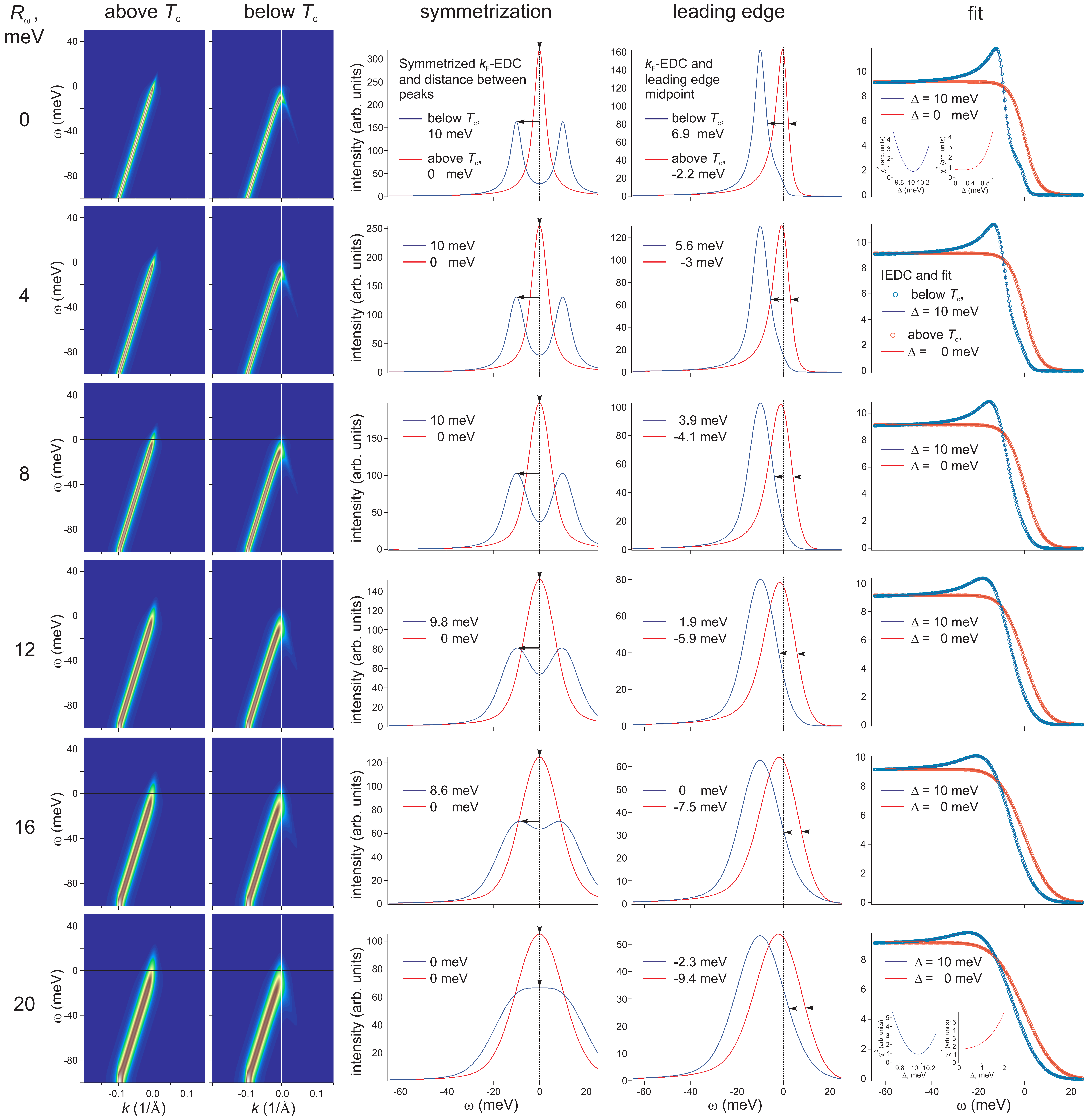}\caption{Influence of the energy resolution on the determination of the gap via symmetrization, leading edge, and fit. First column: energy resolution for the corresponding row. Second column: simulated energy-momentum cut above $T_{\rm c}$ ($\Sigma^{\prime\prime}=3$\,meV, $kT=$3\,meV, $\Delta=0$\,meV). Third column: simulated energy-momentum cut below $T_{\rm c}$ ($\Sigma^{\prime\prime}=3$\,meV, $kT=$1\,meV, $\Delta=10$\,meV). Fourth column: determination of the gap with ``symmetrization''. Fifth column: determination of the gap with ``leading edge''. Sixth column: determination of the gap with fit to formula (1) and $\chi^2$ criterion as insets to some panels. ``Symmetrization'' and ``leading edge'' provide acceptable results for good resolution, and fail when the resolution becomes worse. The fitting procedure always provides the correct result.}
\vspace{-0ex}
\label{f:Model5}
\end{figure*}

\begin{table}[]
\begin{tabular}{l@{~~~}c@{~~~}c@{~~~}l@{~~~}}
\toprule
$R_\omega$  &``Symmetri- &Leading &Fit\\
& zation''&edge shift&\\
\midrule
\phantom{1}0\phantom{.1} &10\phantom{.1}&9.1&$10\pm0.1$\\
\phantom{1}4\phantom{.1} &10\phantom{.1}&8.6&10\\
\phantom{1}8\phantom{.1} &10\phantom{.1}&8.0&10\\
12 &\phantom{1}9.8&7.8&10\\
16 &\phantom{1}8.6&7.5&10\\
20 &\phantom{1}0\phantom{.1}&7.1&$10\pm0.1$\\
\bottomrule
\end{tabular}
\caption{Superconducting gap, as extracted from modeled data (Fig.\,8) with different methods. All numbers are given in millielectron-volts. The correct value of the gap (implemented in simulation) equals 10\,meV.}
\label{tab1}
\end{table}

\subsection{Momentum resolution}
Next, we consider the influence of the experimental momentum resolution on the determination of the gap from ARPES data with ``symmetrization'', ``leading edge'', and the fitting to formula (1). The results of these studies are shown in Fig.\,9 and summarized in Table~IV. Note that \emph{not} the resolution of the analyzer is important, but the resolution of the \emph{whole} photoemission experiment.

The width (full width at half maximum) of the narrowest EDC from Refs.\,2--6, as well as from our studies is 8--10\,meV. The momentum resolution is about 0.1\,$\text{\AA}^{-1}$.

\begin{figure*}[]
\vspace{+0ex}\includegraphics[width=1\textwidth]{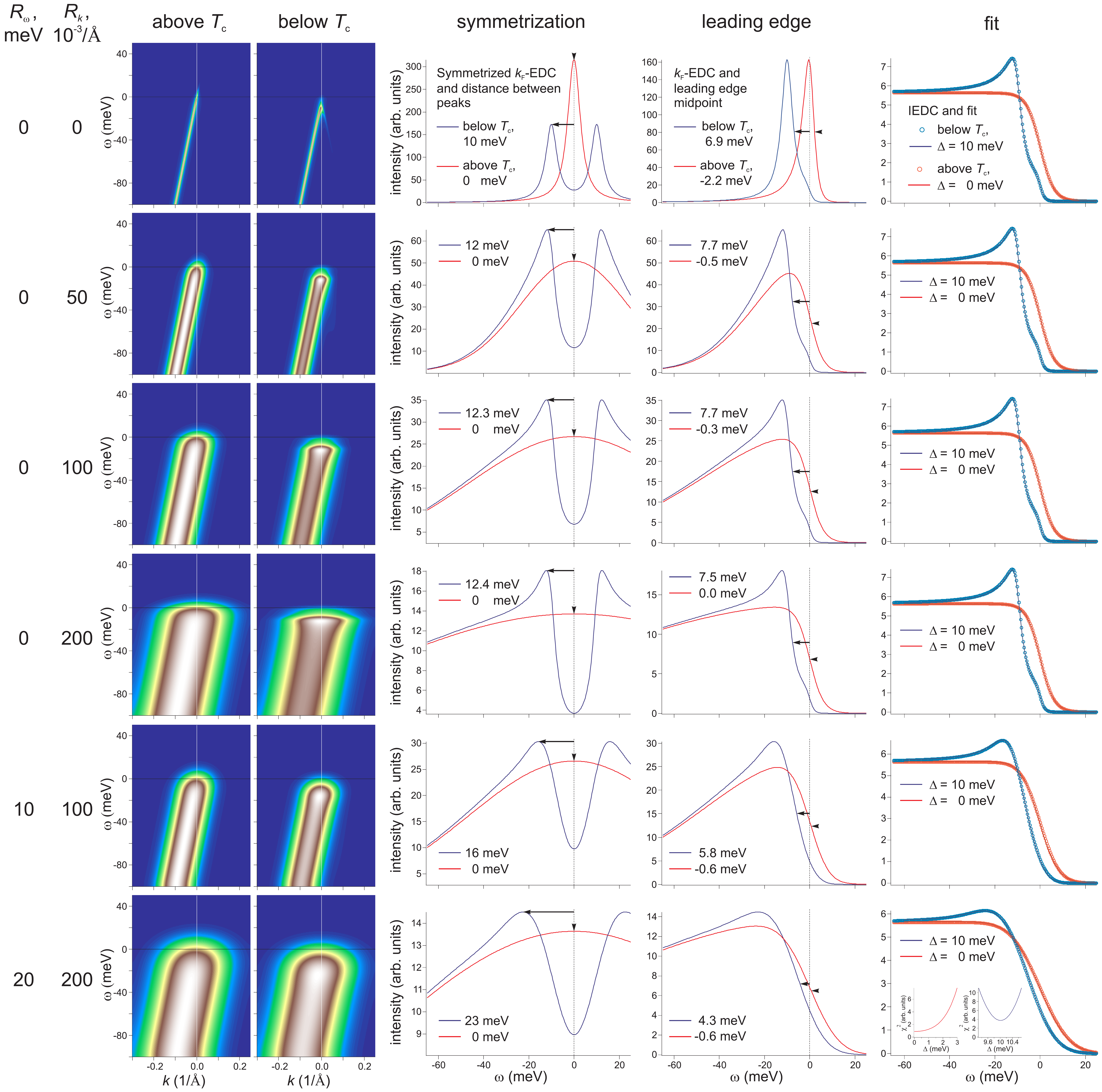}\caption{Influence of the momentum resolution on the determination of the gap via symmetrization, leading edge, and fit. First column: energy resolution for the corresponding row. Second column: momentum resolution for the corresponding row. Third column: simulated energy-momentum cut above $T_{\rm c}$ ($\Sigma^{\prime\prime}=3$\,meV, $kT=3$\,meV, $\Delta=0$\,meV). Fourth column: simulated energy-momentum cut below $T_{\rm c}$ ($\Sigma^{\prime\prime}=3$\,meV, $kT=1$\,meV, $\Delta=10$\,meV). Fifth column: determination of the gap with ``symmetrization''. Sixth column: determination of the gap with ``leading edge''. Seventh column: determination of the gap with fit to formula (1) and $\chi^2$ criterion as insets to some panels. ``Symmetrization'' and ``leading edge'' provide acceptable results for good resolution, and fail when the resolution becomes worse. The fitting procedure always provides the correct result.}
\vspace{-0ex}
\label{f:Model6}
\end{figure*}

\begin{table}[]
\begin{tabular}{l@{~~~}l@{~~~}c@{~~~}c@{~~~}l@{~~~}}
\toprule
$R_\omega$  &$R_k$,&``Symmetri- &Leading &Fit\\
&$10^{-3}\text{\AA}^{-1}$ &zation''&edge shift&\\
\midrule
\phantom{1}0 &\phantom{10}0&10.0&9.1&$10$\\
\phantom{1}0 &\phantom{1}50&12.0&8.2&10\\
\phantom{1}0 &100&12.3&8.0&10\\
\phantom{1}0 &200&12.4&7.5&10\\
\textbf{10} &\textbf{100}&\textbf{16}\phantom{.0}&\textbf{6.4}&\textbf{10}\\
20 &200&23\phantom{.0}&4.9&$10\pm0.5$\\
\bottomrule
\end{tabular}
\caption{Superconducting gap, as extracted from modeled data (Fig.\,9) with different methods. All numbers, except for momentum resolution, are given in millielectron-volts. The correct value of the gap (implemented in simulation) equals 10\,meV. Parameters $R_\omega=10$\,meV and $R_k=0.1\text{\,\AA}^{-1}$ correspond to the experimentally observed widths of the spectra.}
\label{tab1}
\end{table}

\subsection{Nonlinearity of the band dispersion}

The case when the band depth is comparable to the value of the superconducting gap is quite complicated, and really requires special treatment. That is why formula (2) has been derived and used to fit the data. It is easy to mistake the van Hove singularity for the gap when using simplified methods of data analysis. Masking effects of van Hove singularity is one of real examples where ``symmetrization'' and ``leading edge'' give wrong results (Fig.\,10, and especially column three, bottom row).  Naturally, such ``gap'' will not close at $T_{\rm c}$.

Here we have modeled the influence of the nonlinearity of the band dispersion in conjunction with experimental momentum resolution on the determination of the gap from ARPES data with ``symmetrization'', ``leading edge'', and fit to formula (2). The results of these studies are shown in Fig.\,10 and summarized in Table~V.

\begin{figure*}[t!]
\vspace{+0ex}\includegraphics[width=1\textwidth]{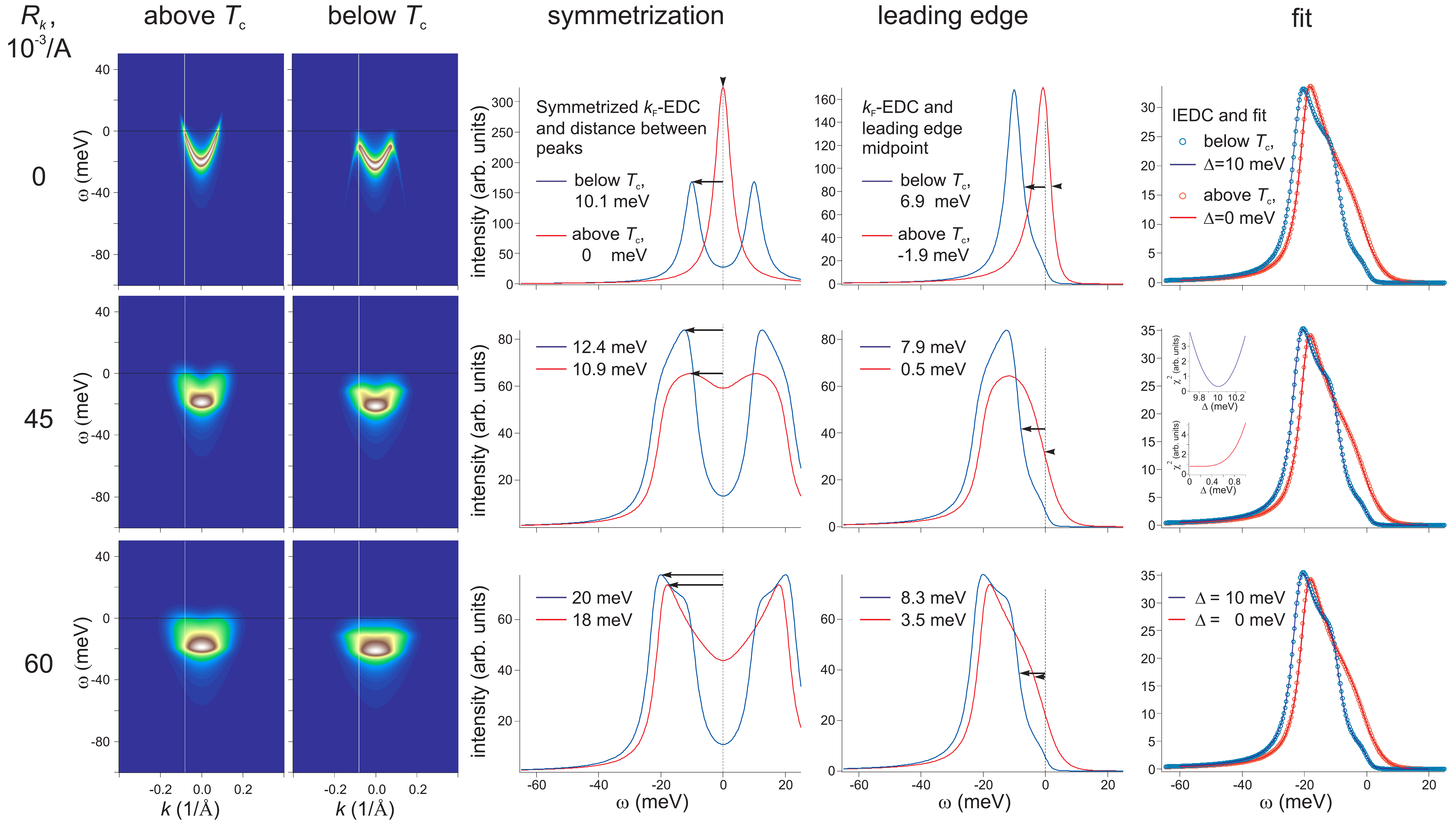}\caption{Influence of the small band depth on the determination of the gap via symmetrization, leading edge, and fit. First column: momentum resolution for the corresponding row. Second column: simulated energy-momentum cut above $T_{\rm c}$ ($\Sigma^{\prime\prime}=3$\,meV, $kT=3$\,meV, $\Delta=0$\,meV, $\varepsilon_0=20$\,meV). Third column: simulated energy-momentum cut below $T_{\rm c}$ ($\Sigma^{\prime\prime}=3$\,meV, $kT=1$\,meV, $\Delta=10$\,meV, $\varepsilon_0=20$\,meV). Fourth column: determination of the gap with ``symmetrization''. Fifth column: determination of the gap with ``leading edge''. Sixth column: determination of the gap with fit to formula (2) and $\chi^2$ criterion as insets to some panels. First row: no resolution effects added. Second row: small resolution effects are added. Third row: moderate resolution effects are added (resolution effects are comparable, and may be even smaller to those in real data, which is easy to see comparing these energy-momentum cuts to directly measured). For simplicity, only momentum resolution is added. ``Symmetrization'' and ``leading edge'' provide acceptable results for good resolution, and fail when the resolution becomes worse. The fitting procedure always provides the correct result.}
\vspace{-0cm}
\label{f:Model4}
\end{figure*}

\begin{table}[]
\begin{tabular}{l@{~~~}c@{~~~}c@{~~~}l@{~~~}l@{~~~}}
\toprule
$R_k$,& Input to & ``Symmetri- & Leading &Fit\\
$10^{-3}\text{\AA}^{-1}$& the model&zation'' & edge &\\
\midrule
\phantom{4}0 &\phantom{1}0&\phantom{1}0\phantom{.1}&$-1.9$&\phantom{1}$0+0.8$\\
 &10&10.1&\hspace{0.24cm}6.9&$10\pm0.1$\\
\midrule
45 &\phantom{1}0&10.9&\hspace{0.24cm}0.5&\phantom{1}$0+0.8$\\
 &10&12.4&\hspace{0.24cm}7.9&$10\pm0.1$\\
\midrule
60 &\phantom{1}0&18\phantom{.1}&\hspace{0.24cm}3.5&\phantom{1}$0+0.8$\\
 &10&20\phantom{.1}&\hspace{0.24cm}8.3&$10\pm0.1$\\
\bottomrule
\end{tabular}
\caption{Superconducting gap, as extracted from modeled data (Fig.\,10) with different methods. All numbers, except for momentum resolution, are given in millielectron-volts.}
\label{tab1}
\end{table}

\subsection{Non-superconducting component}
According to recent $\mu$SR (muon spin rotation) studies, superconducting fraction for optimally doped Ba$_{1-x}$K$_{x}$Fe$_2$As$_2$ samples (those used in Refs.\,2--6) comprises 50\% of the sample volume (see Ref.\,\onlinecite{Uemura}), and for our slightly underdoped samples it comprises only 25\% (see Ref.\,\onlinecite{Hinkov}). Under these circumstances, ``leading edge'' is completely irrelevant to the gap value, while ``symmetrization'' may provide some estimates of the value of the gap depending on other conditions (resolution, lifetime \textit{etc.}). Fitting in this case is indispensable, as it not only reveals precise values of the gap, but also allows to determine the fractions of the superconducting and non-superconducting signals (see Fig.\,5). For our crystals these fractions determined from two completely different methods\,---\,$\mu$SR and the fit of the ARPES data\,---\,perfectly match each other.

\subsection{Renormalization}

Presence of the dispersion anomalies, ``kinks'' can affect position of the leading edge and peaks in the symmetrized EDC, and can be mistaken for energy gap, similarly to the discussed above van Hove singularity. In the case of linear bare band dispersion \cite{bare}, IEDC is not affected by self energy at all:
\begin{equation}
\int^{+\infty}_{-\infty}\frac{1}{2\pi}\frac{\Sigma^{\prime\prime}(\omega)}{(\omega - \Sigma^{\prime}(\omega) - v_{\mathrm{F}}k)^2+{\Sigma^{\prime\prime}}(\omega)^2}\mathrm{d}k = \frac{1}{v_{\mathrm{F}}}.
\end{equation}
\subsection{Summary}

\begin{enumerate}
     \item The proposed fitting procedure is rigorous and precise method of gap extraction, which accounts for several important features of Ba$_{1-x}$K$_{x}$Fe$_2$As$_2$ photoemission spectra:

    \begin{enumerate}

        \item nonlinearity of the band dispersion;

        \item presence of large non-superconducting component;

        \item experimental resolution.

    \end{enumerate}

    \item ``Symmetrization'' is not a universal way for the extraction of the gap from spectroscopic data, since it is highly sensitive to experimental resolution, and non-linearity of the band dispersion. For example, it

    \begin{enumerate}
        \item gives zero value for the gap while there is substantial gap (Fig.\,8, bottom row);

        \item gives substantial value for the gap, while actual gap is zero (Fig.\,10, bottom row).
    \end{enumerate}

    \item ``Leading edge'' alone is not a good measure of the gap (see bottom rows of the Figs.\,8--10), while leading edge \emph{shift} in absence of the non-superconducting component is a quite good, although still rough measure of the gap, and provides result with an accuracy better than 50\% even under severe conditions (see Tables III--V).

\end{enumerate}


\newpage


\begin{thebibliography}{00}
\bibitem{Wang} Cao Wang, Linjun Li, Shun Chi, Zengwei Zhu, Zhi Ren, Yuke Li, Yuetao Wang, Xiao Lin, Yongkang Luo, Shuai Jiang, Xiangfan Xu, Guanghan Cao and Zhu'an Xu, Europhys. Lett. \textbf{83}, 67006 (2008)
\bibitem{Kaminski1} C.\,Liu, T.\,Kondo, M.\,E.\,Tillman, R.\,Gordon, G.\,D.\,Samolyuk, Y.\,Lee, C.\,Martin, J.\,L.\,McChesney, S.\,Bud'ko, M.\,A.\,Tanatar, E.\,Rotenberg, P.\,C.\,Canfield, R.\,Prozorov, B.\,N.\,Harmon, A.\,Kaminski, arXiv:0806.2147 (2008)
\bibitem{Ding}  H.\,Ding, P.\,Richard, K.\,Nakayama, K.\,Sugawara, T.\,Arakane, Y.\,Sekiba, A.\,Takayama, S.\,Souma, T.\,Sato, T.\,Takahashi, Z.\,Wang, X.\,Dai, Z.\,Fang, G.\,F.\,Chen, J.\,L.\,Luo and N.\,L.\,Wang, Europhys.\,Lett. \textbf{83}, 47001 (2008)
\bibitem{Zhou} Lin Zhao, Haiyun Liu, Wentao Zhang, Jianqiao Meng, Xiaowen Jia, Guodong Liu, Xiaoli Dong, G. F. Chen, J. L. Luo, N. L. Wang, Wei Lu, Guiling Wang, Yong Zhou, Yong Zhu, Xiaoyang Wang, Zhongxian Zhao, Zuyan Xu, Chuangtian Chen, X. J. Zhou, Chinese Phys. Lett. \textbf{25}, 4402 (2008)
\bibitem{Kaminski} Takeshi Kondo, A.\,F.\,Santander-Syro, O.\,Copie, Chang Liu, M.\,E.\,Tillman, E.\,D.\,Mun, J.\,Schmalian, S.\,L.\,Bud'ko, M.\,A.\,Tanatar, P.\,C.\,Canfield, A.\,Kaminski, 	 \prl \textbf{101}, 147003 (2008)
\bibitem{Hasan} L.\,Wray, D.\,Qian, D.\,Hsieh, Y.\,Xia, L.\,Li, J.\,G.\,Checkelsky, A.\,Pasupathy, K.\,K.\,Gomes, A.\,V.\,Fedorov, G.\,F.\,Chen, J.\,L.\,Luo, A.\,Yazdani, N.\,P.\,Ong, N.\,L.\,Wang, M.\,Z.\,Hasan, arXiv:0808.2185 (2008)
\bibitem{Chen} T.\,Y.\,Chen, Z.\,Tesanovic, R.\,H.\,Liu, X.\,H.\,Chen and C.\,L.\,Chien, Nature\,(London) \textbf{453}, 1224 (2008)
\bibitem{Szabo} P.\,Szabo, Z.\,Pribulova, G.\,Pristas, S.\,L.\,Bud'ko, P.\,C.\,Canfield, P.\,Samuely, arXiv:0809.1566 (2008)
\bibitem{Mu} Gang Mu, Huiqian Luo, Zhaosheng Wang, Lei Shan, Cong Ren, Hai-Hu Wen, 	arXiv:0808.2941 (2008)
\bibitem{Crystals} G.\,L.\,Sun, D.\,L.\,Sun, M. Konuma, P. Popovich, A. Boris, J. B. Peng, K.-Y. Choi, P. Lemmens and C.\,T.\,Lin, arXiv:0901.2728 (2009)
\bibitem{Volodya} V.\,B.\,Zabolotnyy, D.\,S.\,Inosov, D.\,V.\,Evtushinsky, A.\,Koitzsch, A.\,A.\,Kordyuk, J.\,T.\,Park, D.\,Haug, V.\,Hinkov, A.\,V.\,Boris, D.\,L.\,Sun, G.\,L.\,Sun, C.\,T.\,Lin, B.\,Büchner, A.\,Varykhalov, R.\,Follath, S.\,V.\,Borisenko, Nature \textbf{457}, 569 (2009)
\bibitem{blade} This pocket is named ``blade'' because whole structure around the X-point\,---\,X-pocket and four blades\,---\,reminds a propeller.
\bibitem{Dynes} R.\,C.\,Dynes, V.\,Narayanamurti, and J.\,P.\,Garno, \prl \textbf{41}, 1509 (1978)
\bibitem{Kiss} S.\,Tsuda, T.\,Yokoya, T.\,Kiss, Y.\,Takano, K.\,Togano, H.\,Kito, H.\,Ihara, and S.\,Shin, \prl \textbf{87}, 177006 (2001)
\bibitem{nonSC} Note that non-superconducting part, shown in Fig.\,5(a) also contains momentum-independent part, that should be subtracted.
\bibitem{Hinkov} J.\,T.\,Park, D.\,S.\,Inosov, Ch.\,Niedermayer, G.\,L.\,Sun, D.\,Haug, N.\,B.\,Christensen, R.\,Dinnebier, A.\,V.\,Boris, A.\,J.\,Drew, L.\,Schulz, T.\,Shapoval, U.\,Wolff, V.\,Neu, Xiaoping\,Yang, C.\,T.\,Lin, B.\,Keimer, V.\,Hinkov, arXiv:0811.2224 (2008)
\bibitem{Uemura} T.\,Goko, A.\,A.\,Aczel, E.\,Baggio-Saitovitch, S.\,L.\,Budko, P.\,C.\,Canfield, J.\,P.\,Carlo, G.\,F.\,Chen, Pengcheng Dai, W.\,Z.\,Hu, H.\,Kageyama, G.\,M.\,Luke, J.\,L.\,Luo, N.\,Ni\, D.\,Reznik, D.\,R.\,Sanchez-Candela, A.\,T.\,Savici, K.\,J.\,Sikes, N.\,L.\,Wang, C.\,R.\,Wiebe, T.\,J.\,Williams, T.\,Yamamoto, W.\,Yu, Y.\,J.\,Uemura, arXiv:0808.1425 (2008)
\bibitem{Voigt} D.\,V.\,Evtushinsky, A.\,A.\,Kordyuk, S.\,V.\,Borisenko, V.\,B.\,Zabolotnyy, M.\,Knupfer, J.\,Fink, B.\,Büchner, A.\,V.\,Pan, A.\,Erb, C.\,T.\,Lin, and H.\,Berger, \prb \textbf{74}, 172509 (2006)
\bibitem{Mazin} I.\,I.\,Mazin, M.\,D.\,Johannes, Nature Physics, doi:10.1038/nphys1160 (2009)
\bibitem{kord_LEG} A.\,A.\,Kordyuk, S.\,V.\,Borisenko, M.\,Knupfer, and J.\,Fink, \prb \textbf{67}, 064504 (2003)
\bibitem{no_matrix_elements} Here we can neglect variations of the matrix elements, as relevant scales in momentum space are rather small.
\bibitem{Mahan} Gerald D. Mahan, \textit{Many particle physics}, Plenum Press (1981)
\bibitem{int_conv} Rigorously speaking, we use the following mathematical fact:
$$\quad \int^{+\infty}_{-\infty} \bigl( f_1 \otimes f_2 \bigr)\,\mathrm{d}x = \int^{+\infty}_{-\infty} f_1\,\mathrm{d}x \cdot \int^{+\infty}_{-\infty} f_2\,\mathrm{d}x,$$
which in our case reads
$$\int\!A(k,\omega)\otimes R_k\,\mathrm{d}k = \int A(k,\omega)\,\mathrm{d}k\cdot \int\!R_k(k)\,\mathrm{d}k$$
\bibitem{Valla} H.-B.\,Yang, J.\,D.\,Rameau, P.\,D.\,Johnson, T.\,Valla, A.\,Tsvelik, and G.\,D.\,Gu, Nature \textbf{456}, 77 (2008).
\bibitem{Shin} T.\,Baba, T.\,Yokoya, S.\,Tsuda, T.\,Kiss, T.\,Shimojima, K.\,Ishizaka, H.\,Takeya, K.\,Hirata, T.\,Watanabe, M.\,Nohara, H.\,Takagi, N.\,Nakai, K.\,Machida, T.\,Togashi, S.\,Watanabe, X.-Y.\,Wang, C.\,T.\,Chen, and S.\,Shin, \prl \textbf{100}, 017003 (2008).
\bibitem{bare} A.\,A.\,Kordyuk, S.\,V.\,Borisenko, A.\,Koitzsch, J.\,Fink, M.\,Knupfer, and H.\,Berger \prb \textbf{71}, 214513 (2005)
\end{thebibliography}
\end{document}